\documentstyle[a4,12pt,epsfig]{article}

\title{
\bf Direct Photons in Nuclear Collisions at FAIR Energies
}

\author{
S.M.~Kiselev\footnote{Talk at the session of Russian Academy of Sciences, ITEP, Moscow, 26 - 30 November 2007}
}

\date{ }

\begin{document}

\maketitle

\begin{flushleft}
\small
\it
Institute for Theoretical and Experimental Physics, Moscow, Russia\\
\end{flushleft}

\begin{abstract}
 
Using the extrapolation of existing data estimations of prompt photon 
production at FAIR energies have been made. At $y=y_{c.m.}$ the rapidity density  
of prompt photons with $p_{t}>$ 1.5 GeV/c per central Au+Au event at 25 AGeV 
is estimated as $\sim 10^{-4}$ . With the planed beam intensity
$10^{9}$ per second and 1\% interaction  probability, for 10\% of most
central events one can expect the prompt photon rate $\sim 10^{2}$ photons
per second.

Direct photons from the hadron scenario of ion collisions generated 
by the Hadron-String-Dynamics~(HSD) transport approach with implemented meson scatterings 
$\pi\rho\rightarrow\pi\gamma, \pi\pi\rightarrow\rho\gamma$
have been analyzed. Photons from short-living resonances 
(e.g. $\omega \rightarrow \pi^{0} \gamma$) decaying during the 
dense phase of the collision should be considered as
direct photons. They contribute significantly in the direct
photon spectrum at $p_{t}=0.5 - 1$ GeV/c. At the FAIR energy 25 AGeV
in Au+Au central collisions the HSD generator predicts, as a lower estimate, 
$\gamma_{direct}/\gamma_{\pi^{0}} \simeq$ 0.5\% in the region $p_{t}=0.5 - 1$ GeV/c.  
At $p_{t}=1.5 - 2$ GeV/c $\gamma_{prompt}/\gamma_{\pi^{0}} \simeq$ 2\%.

Thermal direct photons have been evaluated with the Bjorken Hydro-Dynamics (BHD) model.
The BHD spectra differ strongly from the HSD predictions. 
The direct photon spectrum is very sensitive to the initial temperature
parameter $T_{0}$ of the model. The 10 MeV increase in the $T_{0}$ value leads 
to $\sim$ 2 times higher photon yield. 

\end{abstract}

\newpage


\section{Introduction}
The FAIR (Facility for Antiproton and Ion Research)~\cite{FAIR} accelerators 
will provide heavy ion beams up to Uranium
at beam energies ranging from 2 - 45 AGeV (for Z/A=0.5) and up to 35 AGeV for
Z/A=0.4. The maximum proton beam energy is 90 GeV. The planed ion beam
intensity is $10^9$ per second.

The CBM (Compressed Baryonic Matter)~\cite{CBM} detector will have good possibilities 
for vertex reconstruction, tracking and identification of particles (hadrons, leptons
and photons). Though direct photons are of great interest for the research program
of the CBM experiment, a feasibility study has not been done yet.

During high-energy heavy-ion collisions direct photons, defined as photons
not from particle decays, have very little interaction with the surrounding
medium and are therefore not altered by rescattering. Therefore, they provide
a very interesting probe and convey unique and unperturbed information on
the all stages of the collision (see, e.g. the review~\cite{PhysRep2002}).

On the quark-gluon level the main processes are Compton scattering
$q g \rightarrow q \gamma$ and annihilation $q \bar q \rightarrow g \gamma$.
Photons from initial hard NN collisions are named prompt photons and are the main
source at large $p_{t}$.
On the hadron level the main source of direct photons is meson rescatterings:
$\pi \rho \rightarrow \pi \gamma$, $\pi \pi \rightarrow \rho \gamma$,
$\pi K \rightarrow K^{*} \gamma$, $K \rho \rightarrow K \gamma$, ...
First two channels are most important.

Direct photons from a thermalized quark-gluon or hadron system, if any,
are named thermal photons and are the main source at low $p_{t}$.
Most of theoretical predictions for direct photons assume the
local thermalization, evaluate photon production rates from
the equilibrated quark-gluon or hadron system which then are convoluted
with space-time evolution of the system.

Of cause, during the nucleus-nucleus collision there is a non-equilibrium stage.
It can be described by the transport approach free from the local thermalization 
assumption. However, most of existing transport codes do not include the hadronic 
source (meson rescatterings) of direct photons~\cite{Kiselev}.

Identification of direct photons in heavy-ion collisions is a very difficult 
experimental task, especially at low transverse momentum, because of the large 
background from decay photons, mostly from $\pi^{0}$ decays. First direct photons, 
$\gamma_{direct}/\gamma_{measured} \simeq$ 20\% at $p_{t}>$ 1.5 GeV/c, have been 
extracted by the WA98 collaboration at SPS~\cite{WA98data}. The PHENIX collaboration 
at RHIC using a novel analysis technique decreased a systematic error to
2-3\% and revealed direct photons since $p_{t}>$ 1 GeV/c, 
$\gamma_{direct}/\gamma_{measured} \simeq$ 10\% at
$p_{t}=$ 1 - 2 GeV/c~\cite{PHENIXdata}.

Here we would like to estimate for the FAIR energies:  prompt photons
using the extrapolation of existing $p+p \rightarrow \gamma X$ data and photon 
contribution from hadron sources exploiting as the microscopic 
Hadron-String-Dynamics~(HSD) transport approach~\cite{HSD} and 
the  Bjorken Hydro-Dynamics (BHD)~\cite{BHD}.

\section{Prompt photons}
\label{PromptPhotons}
Existing $p+p(\bar p) \rightarrow \gamma X$ data~\cite{Compelation} cover the
energy range $\sqrt{s}=20-1800$ GeV. Thus, The CBM experiment at the FAIR accelerator
can fill the gap $\sqrt{s}<14$ GeV.

\begin{figure}[h]
\begin{center}
\mbox{\epsfig{file=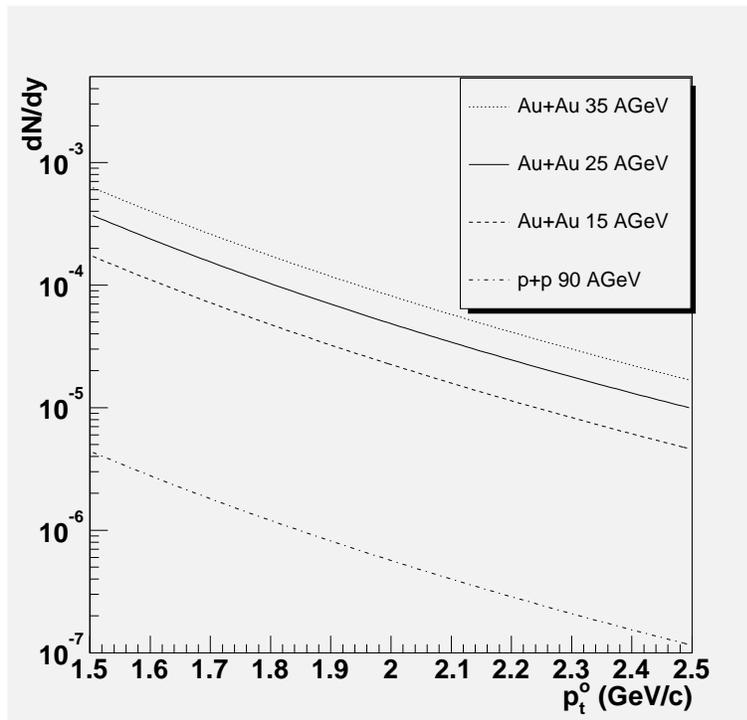,width=10cm}}
\end{center}
\caption{Rapidity density at $y=y_{c.m.}$ of prompt photons with $p_{t}>p_{t}^{0}$ 
at FAIR energies.}
\label{fig:PROMPT}
\end{figure}

At transverse momentum $x_{t}=2p_{t}/\sqrt{s}>0.1$ cross sections can
be fitted in the central rapidity region by the formula
$Ed^{3}\sigma^{pp}/d^{3}p = 575(\sqrt{s})^{3.3}/p_{t}^{9.14}$ 
pb/GeV$^{2}$~\cite{Srivastava}. Using this fit one can estimate the
prompt photon spectrum in nucleus-nucleus, A+B, collisions at the impact
parameter $b$:
$Ed^{3}N^{AB}(b)/d^{3}p = Ed^{3}\sigma^{pp}/d^{3}p \cdot AB \cdot T_{AB}(b)$,
where the nuclear overlapping function is defined as $T_{AB}(b)=N_{coll}(b)/\sigma^{pp}_{in}$,
where $N_{coll}(b)$ is the average number of binary NN collisions.
Nuclear effects (Cronin, shadowing) are ignored in this approach. 
Fig.~\ref{fig:PROMPT} demonstrates extrapolated yield 
of prompt photons at FAIR energies.   
For central Au+Au events ($N_{coll}$=650) at 25 AGeV one can expect $\sim 10^{-4}$
prompt photons with $p_{t}>$ 1.5 GeV/c. At the beam intensity $10^{9}$ per second
and 1\% interaction probability, for 10\% of most central collisions
$\sim 10^{2}$ prompt photons per second are expected.

PYTHIA~\cite{PYTHIA} simulations agree reasonably with the data 
extrapolation to FAIR energies~\cite{Kiselev}. For the prompt photon cross 
section PYTHIA predicts: 
$\sigma \approx 2\cdot10^{-4}$ mb, 85\% from the process $g q \rightarrow \gamma q$ 
and 15\% from $q \bar q \rightarrow \gamma g$.
           
\section{Rescattering and decay photons from HSD}
Cross sections for meson rescatterings $\pi \rho \rightarrow \pi \gamma$ and
$\pi \pi \rightarrow \rho \gamma$ evaluated in the theory of the $\pi$ and $\rho$
meson gas~\cite{Kapusta} have been prepared by the ITEP group~\cite{Kiselev} 
and implemented 
by E.Bratkovskaya into the HSD code. The cross sections diverge at a threshold but 
averaging over the spectral function of the $\rho$ meson solves the problem.

\begin{figure}[tbp]
\begin{center}
\mbox{\epsfig{file=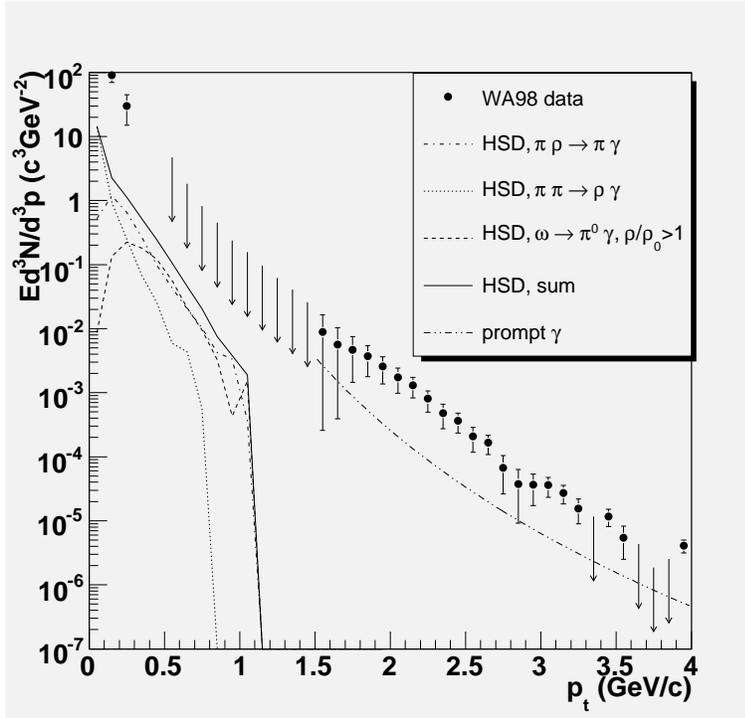,width=10cm}}
\end{center}
\caption{The invariant direct photon spectra for central Pb+Pb events at 158 AGeV in the
central rapidity region $|y-y_{c.m.}|<0.5$.}
\label{fig:WA98data}
\end{figure}

Fig.~\ref{fig:WA98data} shows comparison of HSD predictions with the 
data of the WA98 collaboration at the SPS energy~\cite{WA98data}.
Besides the meson rescatterings, photons from decays of short-living
resonances (e.g. $\omega \rightarrow \pi^{0} \gamma$) are also taken 
into account. In the case when the
life time of a resonance is less than characteristic time of the nucleus-nucleus
collision it is difficult to reconstruct the resonance because the
decay hadron ($\pi^{0}$) can reinteract with surrounding medium especially
if this medium is dense. About 10\% of $\omega \rightarrow \pi^{0} \gamma$ 
decays take place during  the dense nuclear matter stage when 
$\rho/\rho_{0}>1$ ($\rho_{0}$ is the normal nuclear density). 
We will assume that it will not be possible
to reconstruct these decays.

From Fig.~\ref{fig:WA98data} one can see that at $p_{t}=$0.5 - 1 GeV/c
the decay photon contribution is comparable with the contribution from 
$\pi \rho \rightarrow \pi \gamma$. The model prediction
is $\sim$ 10 times lower then one can expect from the experimental data.
There are other sources of direct photons as on the hadron
($K \rho \rightarrow K \gamma$, ...) and the quark-gluon 
($q g \rightarrow q \gamma$, $q \bar q \rightarrow g \gamma$) levels which are
so far not taken into account in the HSD transport code. Thus the HSD results can be
considered as a lower estimate of data.
Photon rapidity densities dN/dy at $y=y_{c.m.}$ are 0.38, 0.26 and 0.17 for
$\pi \rho$, $\pi \pi$ and $\omega$ respectively.
On the same plot the prompt photon estimation ( see the section~\ref{PromptPhotons}) 
is also presented by a line at $p_{t}>$ 1.5 GeV/c.           

\begin{figure}[tbp]
\begin{center}
\mbox{\epsfig{file=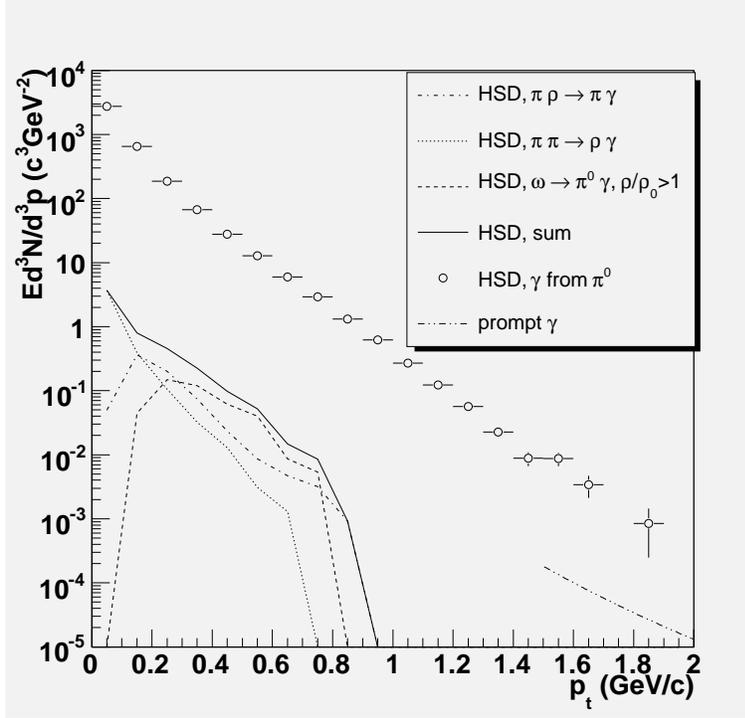,width=10cm}}
\end{center}
\caption{The invariant photon spectra predicted by the HSD generator for 
central (b=0.5 fm) Au+Au events at 25 AGeV in the central rapidity region 
$|y-y_{c.m.}|<0.5$.}
\label{fig:SISInv}
\end{figure}

Fig.~\ref{fig:SISInv} demonstrates predictions of the HSD code at the FAIR energy 
25 AGeV for Au+Au central collisions.
As at the SPS energy here at $p_{t}=$ 0.5 - 1 GeV/c one can observe significant
contribution of photons from the decays  $\omega \rightarrow \pi^{0} \gamma$ in the dense matter. 
On the same plot the photon spectrum from $\pi^{0}$ decays is also shown.
In the region $p_{t}=0.5 - 1$ GeV/c $\gamma_{direct}/\gamma_{\pi^{0}} \simeq$ 0.5\%. 
As have been mentioned it is a lower estimate. 
Rapidity densities dN/dy at $y=y_{c.m.}$ are 0.10, 0.12 and 0.09 for
$\pi \rho$, $\pi \pi$ and $\omega$ respectively.
Average multiplicities per event are 0.31, 0.32 and 0.25 for
$\pi \rho$, $\pi \pi$ and $\omega$ respectively.
At high $p_{t}$=1.5 - 2 GeV/c the part of direct photons is higher
$\gamma_{prompt}/\gamma_{\pi^{0}} \simeq$ 2\%.

\begin{figure}[t]
\begin{center}
\mbox{\epsfig{file=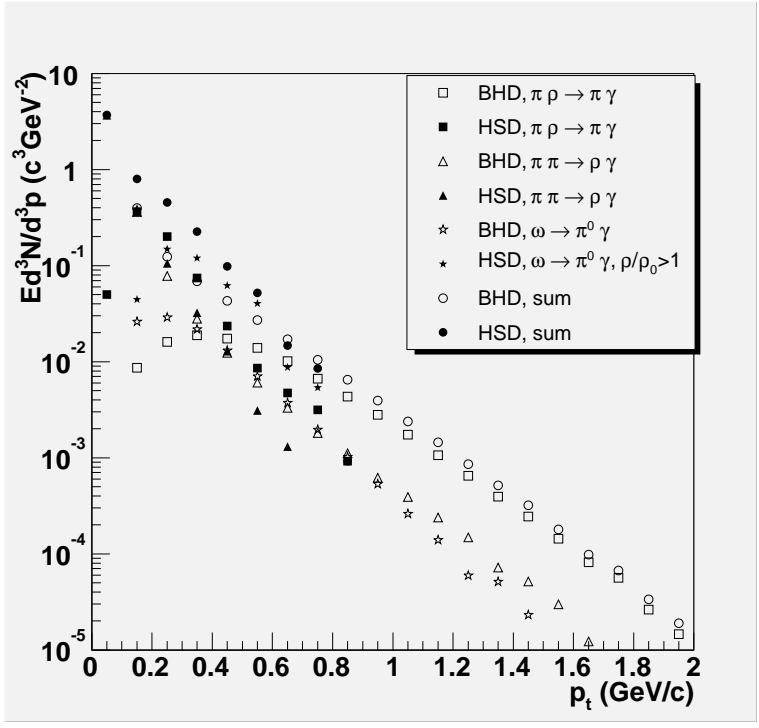,width=10cm}}
\end{center}
\caption{Comparison of the BHD ($\tau_{0}$= 1 fm, $T_{0}$ = 180 MeV) and 
HSD invariant photon spectra for central Au+Au events at 25 AGeV in 
the central rapidity region.}
\label{fig:BHDInv}
\end{figure}
\section{Direct photons from BHD}
In the  Bjorken hydrodynamics~\cite{BHD} it is assumed that 
during the ion-ion collision the system is mainly expanding in beam
direction in a boost-invariant way. Natural variables are
the proper time $\tau=\sqrt{t^{2}-z^{2}}$ and rapidity. 
Thermodynamical variables (pressure, temperature, ...) do not 
depend on the rapidity but are functions of $\tau$, e.g.
$T=T_{0}(\tau_{0}/\tau)^{1/3}$ for an ideal ultrarelativistic gas.
Main initial parameters are the proper time $\tau_{0}$ and temperature $T_{0}$.
Viscosity and conductivity effects are neglected. For this simple space-time
evolution one can evaluate simple formula for photon spectrum
with the photon emission rate as input (the section 2.2.2 of~\cite{PhysRep2002}). 
Photon yield is proportional to $\sim \tau_{0}^{2}$. Though the Bjorken 
scenario is expected to be valid for ultrarelativistic energies (RHIC, LHC)
one can assume it can be used for estimations at SPS and FAIR energies.
The parameterizations presented in~\cite{SongFai} have been used for 
the thermal photon emission rates of the channels $\pi \rho \rightarrow \pi \gamma$ and
$\pi \pi \rightarrow \rho \gamma$. The direct photon WA98 data
can be reproduced at $p_{t}>$ 1.5 GeV/c by the BHD model with the parameters 
$\tau_{0}$= 1 fm and $T_{0}$ = 235 MeV~\cite{kiselevCBM}. 

Fig.~\ref{fig:BHDInv} shows BHD predictions with the parameters $\tau_{0}$= 1 fm, 
$T_{0}$ = 180 MeV at the FAIR energy 25 AGeV for central Au+Au events 
side by side with the HSD results discussed before.
The BHD spectra are broader than ones in the HSD transport approach.
The main reason is the local termalization assumption used in the BHD model.
$\gamma_{direct}/\gamma_{\pi^{0}} \simeq$  0.5\% and 3\% at $p_{t}$= 1 and 1.5 GeV/c 
respectively. 

\begin{figure}[t]
\begin{center}
\mbox{\epsfig{file=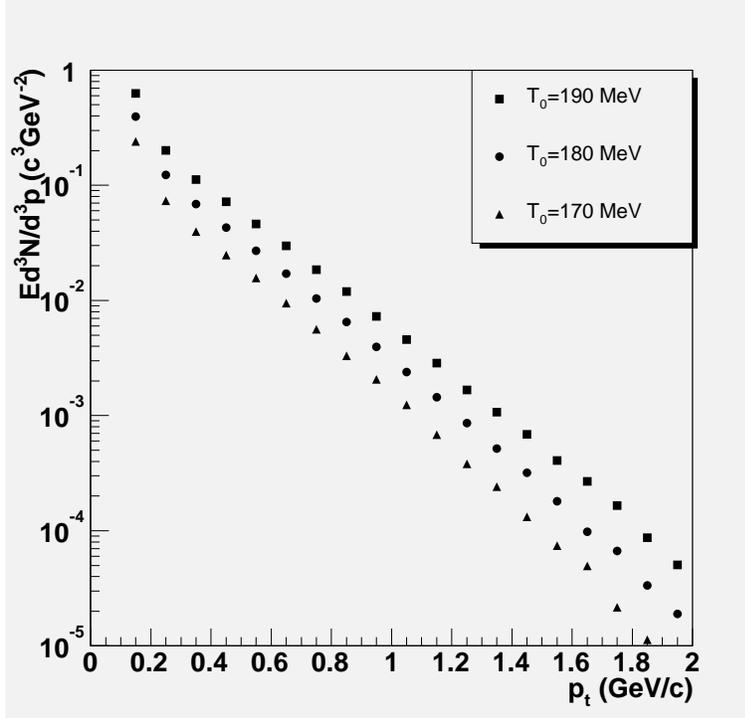,width=10cm}}
\end{center}
\caption{Direct photon spectrum in the
central rapidity regionpredicted by the BHD model
for central Au+Au collisions at 25 AGeV for different values
of the initial parameter $T_{0}$.}
\label{fig:BHDInvT}
\end{figure}
Fig.~\ref{fig:BHDInvT} demonstrates a sensitivity of the direct photon spectrum
to the initial parameter $T_{0}$. The 10 MeV increase in the $T_{0}$ value leads 
to 1.5 and 2.5 times higher photon yield at $p_{t}$= 0.1 GeV/c and 2 GeV/c respectively.


\section{Summary}
Using the extrapolation of existing data estimations of prompt photon 
production at FAIR energies have been made. At $y=y_{c.m.}$ the rapidity density  
of prompt photons with $p_{t}>$ 1.5 GeV/c per central Au+Au event at 25 AGeV 
is estimated as $\sim 10^{-4}$ . With the planed beam intensity
$10^{9}$ per second and 1\% interaction  probability, for 10\% of most
central events one can expect the prompt photon rate $\sim 10^{2}$ photons
per second.

Direct photons from the hadron scenario of ion collisions generated 
by the HSD transport approach with implemented meson scatterings 
$\pi\rho\rightarrow\pi\gamma, \pi\pi\rightarrow\rho\gamma$
have been analyzed. Photons from short-living resonances 
(e.g. $\omega \rightarrow \pi^{0} \gamma$) decaying during the 
dense phase of the collision should be considered as
direct photons. They contribute significantly in the direct
photon spectrum at $p_{t}=$ 0.5 - 1 GeV/c. At the FAIR energy 25 AGeV
in Au+Au central collisions the HSD generator predicts, as a lower estimate, 
$\gamma_{direct}/\gamma_{\pi^{0}} \simeq$ 0.5\% in the region $p_{t}=0.5 - 1$ GeV/c.
At $p_{t}=1.5 - 2$ GeV/c $\gamma_{prompt}/\gamma_{\pi^{0}} \simeq$ 2\%.

Thermal direct photons have been evaluated with the BHD model.
The BHD spectra differ strongly from the HSD predictions. 
The direct photon spectrum is very sensitive to the initial temperature
parameter $T_{0}$ of the model. The 10 MeV increase in the $T_{0}$ value leads 
to $\sim$ 2 times higher photon yield. 

This  work  was partially supported by the Russian Foundation for Basic Research, 
grant number 06-08-01555 and Federal agency of Russia for atomic energy (Rosatom).

\end{document}